# Effect of edge structures on elastic modulus and fracture of graphene nanoribbons under uniaxial tension

**Qiang Lu and Rui Huang**

*Department of Aerospace Engineering and Engineering Mechanics, University of Texas, Austin, Texas 78712, USA*

**ABSTRACT**

Based on atomistic simulations, the nonlinear elastic properties of monolayer graphene nanoribbons under quasistatic uniaxial tension are predicted, emphasizing the effect of edge structures (armchair and zigzag, without and with hydrogen passivation). The results of atomistic simulations are interpreted within a theoretical framework of thermodynamics, which enables determination of the nonlinear functions for the strain-dependent edge energy and the hydrogen absorption energy, for both zigzag and armchair edges. Due to the edge effects, the initial Young's modulus of graphene nanoribbons under infinitesimal strain varies with the edge chirality and the ribbon width. Furthermore, it is found that the nominal strain to fracture is considerably lower for armchair graphene nanoribbons than for zigzag ribbons. Two distinct fracture mechanisms are identified, with homogeneous nucleation for zigzag ribbons and edge-controlled heterogeneous nucleation for armchair ribbons. Hydrogen passivation of the edges is found to have negligible effect on the mechanical properties of zigzag graphene nanoribbons, but its effect is more significant for armchair ribbons.

To harvest the unique physical properties of monolayer graphene for potential applications in nanoelectronics and electromechanical systems, graphene ribbons with nanoscale widths (W < 20 nm) have been produced recently either by lithographic patterning [1-3] or by chemically derived self assembly processes [4]. The edges of the graphene nanoribbons (GNRs) could be zigzag, armchair, or a mixture of both [5]. It has been theoretically predicted that the special characteristics of the edge states leads to a size effect in the electronic state of graphene and controls whether the GNR is metallic, insulating, or semiconducting [5-8]. The effects of the edge structures on deformation and mechanical properties of GNRs have also been studied to some extent [9-18]. On one hand, elastic deformation of GNRs has been suggested as a viable method to tune the electronic structure and transport characteristics in graphene-based devices [15, 16]. On the other hand, plastic deformation and fracture of graphene may pose a fundamental limit for reliability of integrated graphene structures.

The mechanical properties of bulk graphene (infinite lattice without edges) have been studied both theoretically [19-21] and experimentally [22]. For GNRs, however, various edge structures are possible [23, 24], with intricate effects on the mechanical properties. Ideally, the mechanical properties of GNRs may be characterized experimentally by uniaxial tension tests. To date however no such experiment has been reported, although similar tests were performed for carbon nanotubes (CNTs) [25]. Theoretically, previous studies on the mechanical properties of GNRs have largely focused on the linear elastic properties (e.g., Young's modulus and Poisson's ratio) [11-15]. While a few studies have touched upon the nonlinear mechanical behavior including fracture of GNRs [12, 13, 16], the effect of the edge structures in the nonlinear regime has not been well understood. In the present study, based on molecular mechanics (MM) simulations and a thermodynamics theory, we predict the nonlinear elastic behavior of monolayer GNRs under quasistatic uniaxial tension, emphasizing the effect of edge structures in both the linear and nonlinear regimes.

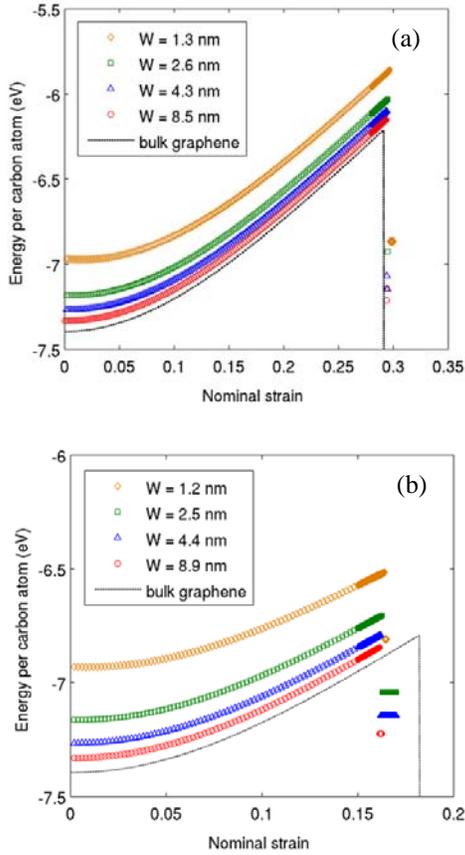

**Figure 1** Potential energy per carbon atom as a function of the nominal strain for graphene nanoribbons under uniaxial tension, with (a) zigzag and (b) armchair edges, both unpassivated. The dashed lines show the results for bulk graphene under uniaxial tension in the zigzag and armchair directions.

Atomistic simulations of GNRs under uniaxial tension are performed using the second-generation reactive empirical bond order (REBO) potential [26]. In each simulation, the tensile strain is applied incrementally in the longitudinal direction of the GNR, until fracture occurs. At each strain level, the statically equilibrium lattice structure of the GNR is calculated to minimize the total potential energy by a quasi-Newton algorithm. Periodic boundary conditions are applied at both ends of the GNR, whereas the two parallel edges (zigzag or armchair) of the GNR are free of external constraint. To study the effect of hydrogen passivation along the free edges, the results for GNRs with bare and passivated edges are compared. The mechanical behavior of infinite graphene lattice is also simulated by applying the periodic boundary conditions at all four edges, for which the uniaxial stress state is achieved by lateral relaxation perpendicular to the loading direction.

Figure 1 shows the results from atomistic simulations for GNRs with unpassivated edges, where the ribbon width ($W$) is varied between 1 and 10 nm. For each GNR, the average potential energy per carbon atom increases as the nominal strain increases until it fractures at a critical strain. To understand the numerical results, we adopt a simple thermodynamics model for the uniaxially stressed GNRs. For a GNR of width $W$ and length $L$, the total potential energy as a function of the nominal strain consists of contributions from deformation of the interior lattice (bulk strain energy) and from the edges (edge energy), namely

$$\phi(\varepsilon) = \phi_0 WL + U(\varepsilon)WL + 2\gamma(\varepsilon)L, \quad (1)$$

where $\varepsilon$ is the nominal strain (relative to the bulk graphene lattice at the ground state), $\phi_0$ is the potential energy density (per unit area) of graphene at the ground state, $U(\varepsilon)$ is the bulk strain energy density (per unit area), and $\gamma(\varepsilon)$ is the edge energy density (per unit length of the edges). While the bulk strain energy density as a function of the nominal strain can be obtained directly from the atomistic calculations for the infinite graphene lattice (dashed lines in Fig. 1), the edge energy density function is determined by subtracting the bulk energy from the total potential energy of the GNRs based on Eq. (1). Thus, both the energy functions are atomistically determined, which can then be fitted with nonlinear polynomial functions for theoretical purposes [20].

The GNR under uniaxial tension is subjected to a net force ($F$) in the longitudinal direction. At each strain increment, the mechanical work done by the longitudinal force equals the increase of the total potential energy, which can be written in a variational form, i.e.,

$$\delta\phi = FL\delta\varepsilon. \quad (2)$$

Consequently, the force ($F$) can be obtained from the derivative of the potential energy function in Eq. (1). A two-dimensional (2-D) nominal stress can then be defined without ambiguity as the force per unit width of the GNR, namely

$$\sigma(\varepsilon) = \frac{F}{W} = \frac{dU}{d\varepsilon} + \frac{2}{W}\frac{d\gamma}{d\varepsilon}. \quad (3)$$

Alternatively, the stress may be calculated directly from the interatomic forces or the virial method [21], where care must be taken for physical interpretation of the atomistic definition of stresses [27].

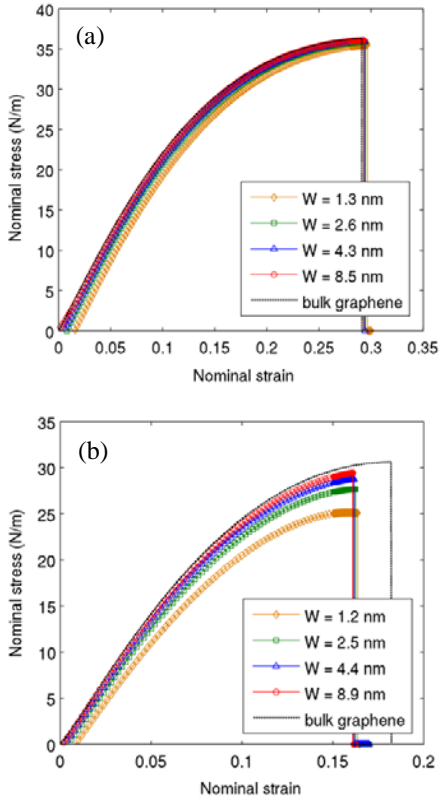

**Figure 2** Nominal stress-strain curves for graphene nanoribbons under uniaxial tension, with (a) zigzag and (b) armchair edges, both unpassivated. The dashed lines show the results for bulk graphene under uniaxial tension in the zigzag and armchair directions.

Figure 2 shows the nominal stress-strain curves of the GNRs obtained by taking the derivative of the potential energy curves in Fig. 1. Apparently, the stress-strain relation is generally nonlinear, for which the tangent modulus as a function of the nominal strain is defined as

$$E(\varepsilon) = \frac{d\sigma}{d\varepsilon} = \frac{d^2U}{d\varepsilon^2} + \frac{2}{W}\frac{d^2\gamma}{d\varepsilon^2}, \quad (4)$$

where the first term on the right-hand side represents the tangent modulus of the bulk graphene lattice (under the condition of uniaxial stress), and the second term is the contribution from the edges (i.e., edge modulus). Therefore, the elastic modulus of the GNR in general depends on the ribbon width ($W$) as well as the edge chirality. Similar stress-strain curves were obtained by molecular dynamics (MD) simulations [12], where the critical strain to fracture is typically lower than the static MM simulations. In addition, Poisson's ratio of the GNRs may also be determined from the same atomistic simulations by calculating the ribbon width as a function of the longitudinal strain, which in general varies nonlinearly with the strain and depends on the edge structure as well.

It is noted from Fig. 2 that the nominal strain in a GNR does not equal zero when the nominal stress is zero. As shown in a previous study [18], this offset strain is inversely proportional to the ribbon width, due to the effect of compressive edge forces for both the zigzag and armchair edges. Recall that the nominal strain is measured relative to the bulk graphene at the ground state. It was also predicted that a fully-relaxed GNR would have periodically buckled edges [9, 10, 18], which in turn would affect the initial stress-strain behavior of the GNR. It is found that the edge buckling of the GNRs essentially disappears under uniaxial tension with the nominal strain beyond a fraction of one percent.

The nominal stress-strain curves in Fig. 2 show approximately linear elastic behavior of all GNRs at relatively small strains (e.g., $\varepsilon < 5\%$). Following Eq. (4), the initial Young's modulus of the GNRs in the linear regime can be written as

$$E_0 = E_0^b + \frac{2}{W}E_0^e, \quad (5)$$

where $E_0^b$ is the initial Young's modulus of the bulk graphene and $E_0^e$ is the initial edge modulus. While the bulk graphene is isotropic in the regime of linear elasticity, the edge modulus depends on the edge chirality with different values for the zigzag and armchair edges. As a result, the initial Young's modulus of the GNR is anisotropic and depends on the ribbon width, as shown in Fig. 3. The REBO potential used in the present study predicts a bulk Young's modulus, $E_0^b = 243$ N/m, and the predicted edge modulus is $E_0^e = 8.33$ nN (~52 eV/nm) for the unpassivated zigzag edge and $E_0^e = 3.65$ nN (~22.8 eV/nm) for the unpassivated armchair edge. With positive moduli for both edges, the Young's modulus of unpassivated GNRs increases as the ribbon width decreases. We note that the predicted edge modulus is considerably lower than a previous calculation using a different potential [11], and the REBO potential is known to underestimate the bulk modulus [28, 29].

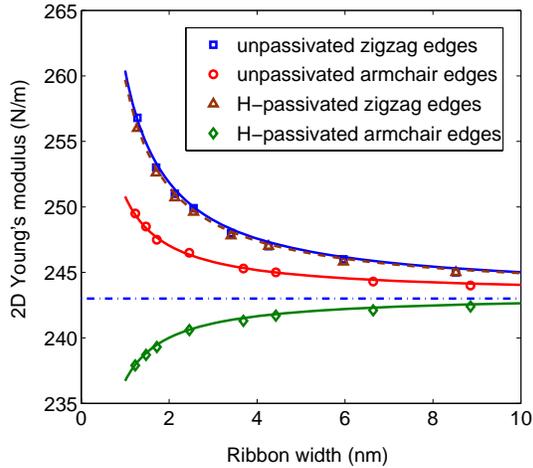

**Figure 3** Initial Young's modulus versus ribbon width for GNRs with unpassivated and hydrogen-passivated edges. The horizontal dot-dashed line indicates the initial Young's modulus of bulk graphene predicted by the REBO potential.

For a GNR with hydrogen (H) passivated edges, the potential energy in Eq. (1) is modified to account for the hydrogen adsorption, namely

$$\phi(\varepsilon) = \phi_0 WL + U(\varepsilon)WL + 2\gamma(\varepsilon)L - 2\gamma_H(\varepsilon)L, \quad (6)$$

where $\gamma_H(\varepsilon)$ is the adsorption energy of hydrogen per unit length along the edges of the GNR and the negative sign indicates typically reduced edge energy due to hydrogen passivation [17, 23]. By comparing the potential energies for the GNRs with and without H-passivation, the adsorption energy can be determined as a function of the nominal strain for both armchair and zigzag edges. At zero strain ($\varepsilon = 0$), our MM calculations predict the hydrogen adsorption energy to be 20.5 and 22.6 eV/nm for the zigzag and armchair edges, respectively, which compare closely with the first-principle calculations [23]. Under uniaxial tension, the adsorption energy varies with the nominal strain. Similar to Eq. (3), the nominal stress for the H-passivated GNR is obtained as

$$\sigma(\varepsilon) = \frac{dU}{d\varepsilon} + \frac{2}{W}\left(\frac{d\gamma}{d\varepsilon} - \frac{d\gamma_H}{d\varepsilon}\right). \quad (7)$$

Correspondingly, the tangent modulus is

$$E(\varepsilon) = \frac{d\sigma}{d\varepsilon} = \frac{d^2U}{d\varepsilon^2} + \frac{2}{W}\left(\frac{d^2\gamma}{d\varepsilon^2} - \frac{d^2\gamma_H}{d\varepsilon^2}\right). \quad (8)$$

The effect of hydrogen passivation on the initial Young's modulus of GNRs is shown in Fig. 3. Interestingly, while hydrogen passivation has negligible effect on the initial Young's modulus of GNRs with zigzag edges, the effect is dramatic for GNRs with armchair edges. In the latter case, a negative edge modulus ($E_0^e = -20.5$ eV/nm) is obtained, and thus the initial Young's modulus decreases with decreasing ribbon width, opposite to the unpassivated GNRs. A previous study [11] using a different potential predicted a negative edge modulus for the H-passivated zigzag edge and a positive edge modulus for the passivated armchair edge. The discrepancy highlights the uncertainty in the quantitative predictions made by the empirical potentials. It is thus necessary to develop alternative potentials to improve the accuracy of atomistic calculations [30, 31].

Without any defect, the bulk graphene fractures when the tangent modulus becomes zero (i.e., $d^2U/d\varepsilon^2 = 0$), dictated by the intrinsic lattice instability under tension [19, 21, 32]. At a finite temperature, however, fracture may occur much earlier due to thermally activated processes [12]. As shown in a previous study [21], the critical strain to fracture for bulk graphene varies with the loading direction, because the hexagonal lattice of graphene preferably fractures along the zigzag directions by cleavage. As shown in Fig. 1a, the GNRs with zigzag edges fracture at a critical strain close to that of bulk graphene loaded in the same direction. In contrast,

Fig. 1b shows that the GNRs with armchair edges fracture at a critical strain considerably lower than bulk graphene. In both cases, the fracture strain slightly depends on the ribbon width, as shown in Fig. 4. Hydrogen passivation of the edges leads to slightly lower fracture strains for zigzag GNRs, but slightly higher fracture strains for armchair GNRs.

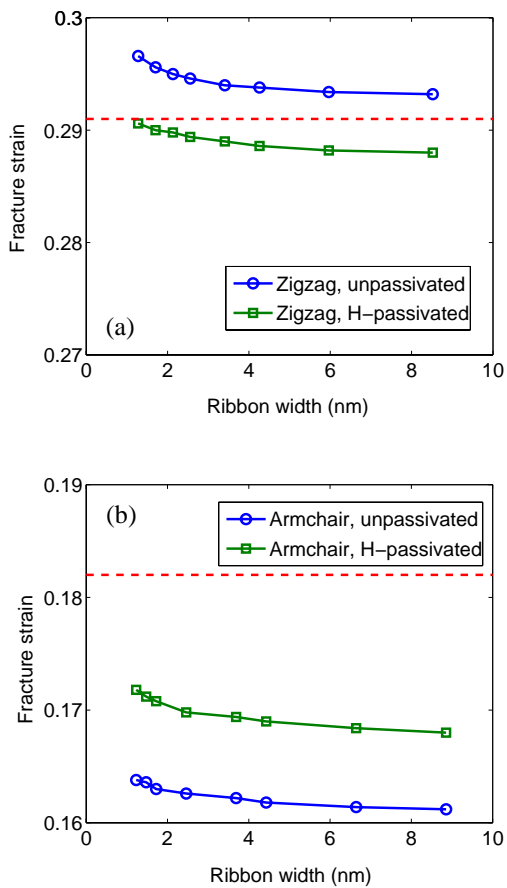

**Figure 4** Fracture strain versus ribbon width for GNRs under uniaxial tension, with (a) zigzag and (b) armchair edges. The horizontal dashed line in each figure indicates the fracture strain of bulk graphene under uniaxial tension in the same direction.

The processes of fracture nucleation in GNRs are studied by molecular dynamics (MD) simulations at different temperatures. It is found that the edge effect leads to two distinct mechanisms for fracture nucleation in GNRs at relatively low temperatures ($T < 300$ K). Figure 5 shows two fractured GNRs at 50 K. For the GNR with zigzag edges (Fig. 5a), fracture nucleation occurs stochastically at the interior lattice of the zigzag GNRs. As a result, the fracture strain is very close to that of bulk graphene strained in the same direction. However, for the GNR with armchair edges (Fig. 5b), fracture nucleation occurs exclusively near the edges. Thus, the armchair edge serves as the preferred location for fracture nucleation, leading to a lower fracture strain compared to bulk graphene. Two distinct fracture nucleation mechanisms are thus identified as interior homogeneous nucleation for the zigzag GNRs and edge-controlled heterogeneous nucleation for the armchair GNRs. The same mechanisms hold for GNRs with H-passivated edges. It is evident from Fig. 5 that cracks preferably grow along the zigzag directions of the graphene lattice in both cases. Formation of suspended atomic chains is often observable from the MD simulations as shown in Fig. 5. Similar chain formation was observed in experiments [33] and a first-principle study [16].

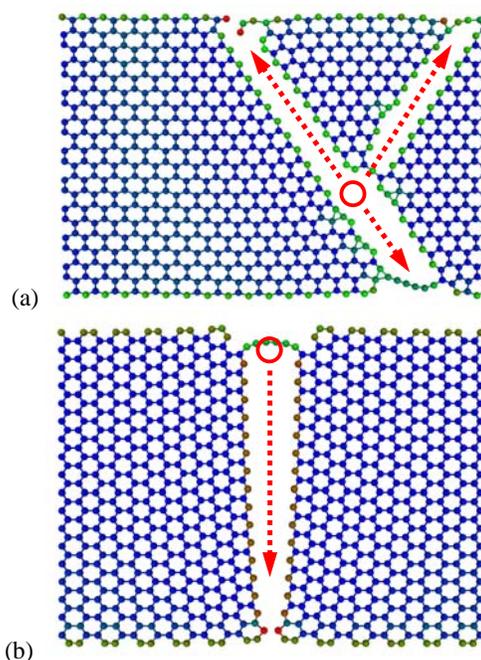

**Figure 5** Fracture of graphene nanoribbons under uniaxial tension. (a) Homogeneous nucleation for a zigzag GNR; (b) edge-controlled heterogeneous nucleation for a armchair GNR. The circles indicate the nucleation sites, and the arrows indicate the directions of crack growth. Color indicates the potential energy of the carbon atoms.

In addition to the fracture strain, the nominal fracture stress (uniaxial tensile strength) of the GNRs can be determined from the stress-strain curves in Fig. 2. As shown in Fig. 6, the fracture stress decreases for GNRs with unpassivated edges.

Hydrogen passivation of the edges slightly increases the fracture stress. The edge effect is relatively small for zigzag GNRs, with all the fracture stresses around 36 N/m, very close to that of bulk graphene. For the armchair GNRs, the fracture stress can be considerably lower, e.g., 27.5 N/m for an unpassivated GNR with $W$ = 2.5 nm, comparing to 30.6 N/m for bulk graphene under uniaxial tension in the armchair direction.

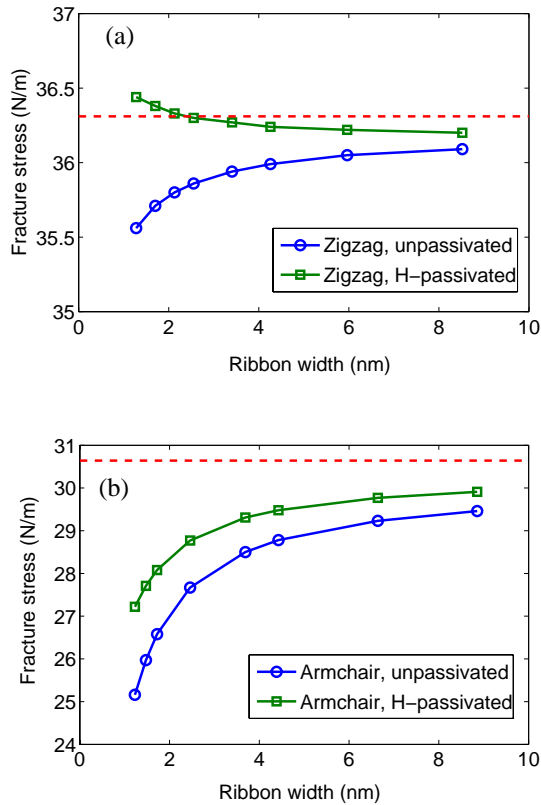

**Figure 6** Nominal fracture stress versus ribbon width for GNRs under uniaxial tension, with (a) zigzag and (b) armchair edges. The horizontal dashed line in each figure indicates the fracture stress of bulk graphene under uniaxial tension in the same direction.

In summary, this paper presents a theoretical study on the effects of edge structures on the mechanical properties of graphene nanoribbons under uniaxial tension. Due to the edge effect, the initial Young's modulus of GNRs under infinitesimal strain depends on both the chirality and the ribbon width. Furthermore, it is found that the strain to fracture is considerably lower for armchair GNRs than that for zigzag GNRs. Two distinct fracture mechanisms are identified, with homogeneous nucleation for the zigzag GNRs and edge-controlled heterogeneous nucleation for the armchair GNRs. Hydrogen passivation is found to have relatively small effects on the mechanical behavior of zigzag graphene ribbons, but its effect is more significant for armchair ribbons.

## Acknowledgements
The authors gratefully acknowledge funding of this work by National Science Foundation through Grant No. 0926851.